# Interplay of phonon directionality and emission polarization in two-dimensional layered metal halide perovskites

*Roman Krahne[1], Miao-Ling Lin[2,3], and Ping-Heng Tan[2,3]*

[1] Optoelectronics, Istituto Italiano di Tecnologia (IIT), Via Morego 30, 16163 Genoa, Italy

[2] State Key Laboratory of Superlattices and Microstructures, Institute of Semiconductors, Chinese Academy of Sciences, 100083 Beijing, China

[3] Center of Materials Science and Optoelectronics Engineering & CAS Center of Excellence in Topological Quantum Computation, University of Chinese Academy of Sciences, 100049 Beijing, China.

CONSPECTUS

Layered metal halide perovskites have recently emerged as a highly interesting materials for both basic science and optoelectronic applications. They represent a natural quantum well system for charge carriers that provides rich physics, and the organic encapsulation of the inorganic metal halide layers increases their stability in devices as compared to their 3D counterparts. The large variety of available organic cations enables an immense design freedom to design functional layered perovskite structures. Intriguingly, these organic moieties strongly impact the optical, electrical and mechanical properties not only through their dielectric, elastic, and chemical properties, but also because their accommodation induces mechanical distortions in the inorganic layers due to the soft nature of the crystal lattice of these two-dimensional layered perovskites (2DLPs). Common consensus is that the exciton-phonon coupling plays an important role in the emissive recombination process, however, the detailed mechanism can depend strongly on the material, and often is not clear. For bulk and some 2D materials, the band edge emission line width broadening can be described by the classic models valid for polar inorganic semiconductors, while for the temperature dependence of the self-trapped exciton line width an analysis developed for color centers has been successfully applied. However, for many 2DLPs these approaches do not hold, because their vibrational response consists of a large number of modes which do not feature one dominant LO-phonon. Nevertheless, their emission is strongly determined by electron-phonon coupling, both in terms of the exciton fine structure at the band edge, and self-trapped exciton emission.

With polarized and angle-resolved Raman spectroscopy studies on single 2DLP flakes with different ammonium molecules as organic cations in 2020 we revealed the very rich phonon spectra in the low-frequency regime. Although the phonon bands at low frequency can generally be attributed to the vibrations of the inorganic lattice, we found very different phonon spectra for the same lead-bromide octahedra composition by only changing the type of the organic cations. In addition, the intensity of the different phonon modes depended strongly on the angle of the linearly polarized excitation beam with respect to the in-plane axes of the octahedra lattice. In 2022, we mapped this angular dependence of the phonon modes, which enabled to identify the directionality of the different lattice vibrations. By correlating the phonon spectra with the temperature-dependent emission for a set of 2DLPs that featured very different STE emission, we demonstrated that the exciton relaxation cannot be related to coupling with a single (LO) phonon band, and that several phonon bands should be involved in the emission process. To gain insights into the exciton-phonon coupling effects on the band edge emission, we performed angle-resolved polarized emission and Raman spectroscopy on the same two-dimensional lead iodide perovskite microcrystals. These experiments revealed the impact of the organic cations on the linear polarization of the emission, and corroborated our interpretation that multiple phonon bands should be involved in the radiative recombination process. Analysis of the temperature-dependent line width broadening of the band edge emission showed that for many systems the behavior cannot be described by assuming the involvement of one phonon mode in the electron-phonon coupling process. In summary, our studies revealed a wealth of highly directional low-frequency phonons in 2DLPs from which several bands are involved in the emission process, which leads to diverse optical and vibrational properties depending on the type of organic cation in the material.

1. **Introduction**

Material science is evolving towards increasingly complex structures that combine the properties of several different components in composites,[1] self-assembled structures of different dimensions,[2] and other architectures. In this respect, hybrid organic-inorganic metal halide perovskites have recently gained strong interest as an extremely versatile platform that combine the optoelectronic properties of solid-state matter with the softness, flexibility and the huge variety of organic molecules.[3] Following their astonishing success in solar cells, these materials proved also highly versatile in light emission, optoelectronics, and other fields.[4] Their architecture combines an inorganic octahedra lattice with semiconductor properties with organic cations that fill the voids between the octahedra and act as a glue to keep the lattice together. This mixed organic-inorganic composition resulted in a wide range of peculiar features both in their optical and elastic properties, for example defect tolerance with respect to their optoelectronic behavior.[5] Low-dimensional metal halide perovskites[6] evolved from the bulk materials by the introduction of organic species that are too large to fit into the voids of the octahedra lattice. With such large organic cations, typically consisting of an amine head group that binds to the metal halide octahedra and an aliphatic chain or aromatic ring as a backbone, layered structures can be achieved that consist or alternating inorganic octahedra lattices and organic layers,[7] as illustrated in Figure 1A. Controlling the number (n) of adjacent octahedra in the vertical direction in the inorganic layer by the ratio of large to small organic cations in the synthesis provides a handle to tune the confinement of the electrical carries in the semiconducting layer from quasi bulk-like (large n) to two-dimensional (2D) with small n.[8] With only one ammonium species, purely 2D crystals (n=1) are obtained that are the subject of this paper, for which we consider the Ruddlesden-Popper phase that features a tail-to-tail connection of the organic cations at the center of the organic layer that leads to a Van-der-Waals gap.[9] This allows for a very large freedom in the choice of the organic molecules,[10] and favors the mechanical exfoliation of thin flakes from larger crystals,[11] as in other 2D materials such as graphene and transition metal chalcogenides. Low-dimensional metal halide perovskites combine inorganic (metal-halide octahedra) lattices that can be described by band structure physics and collective lattice vibrations (phonons) with organic materials such as molecules and polymers whose behavior is governed by molecular orbitals and vibrations of bonds. Building on the extensive knowledge on solid-state 2D materials, their optical and elastic properties are typically described in terms of electronic bands and optical and acoustic phonon modes. [12] [13] In particular, the analysis of the electron-phonon coupling, [14] considering longitudinal-optical (LO) phonons and Fröhlich coupling, has been applied to describe the emission line width broadening in 2D layered perovskites (2DLPs), in analogy as for inorganic (nano) crystal materials. [15-17] [18] However, the vibrational response, or phonon spectrum, of 2DLPs can consists of a very large number of phonon bands, where the identification of a single phonon mode that governs the electron-phonon coupling is not possible. [19,20] Therefore, such analysis can yield at best reasonable average values for the phonon energies and coupling strengths that contribute to the emission broadening. Our studies on the low-frequency vibrational properties and temperature dependent emission of 2DLPs evidenced the limits of such a solid-state analysis of the electron-phonon coupling, [20, 21] which is one focus of the discussion in this account. Furthermore, we will give a comprehensive overview on the directionality of the phonon modes with respect to the axes of the octahedra lattice based on our angle-resolved polarized Raman spectroscopy experiments, where we investigated 2DLPs with different halides, and different metal and organic cations. [20, 22-24]

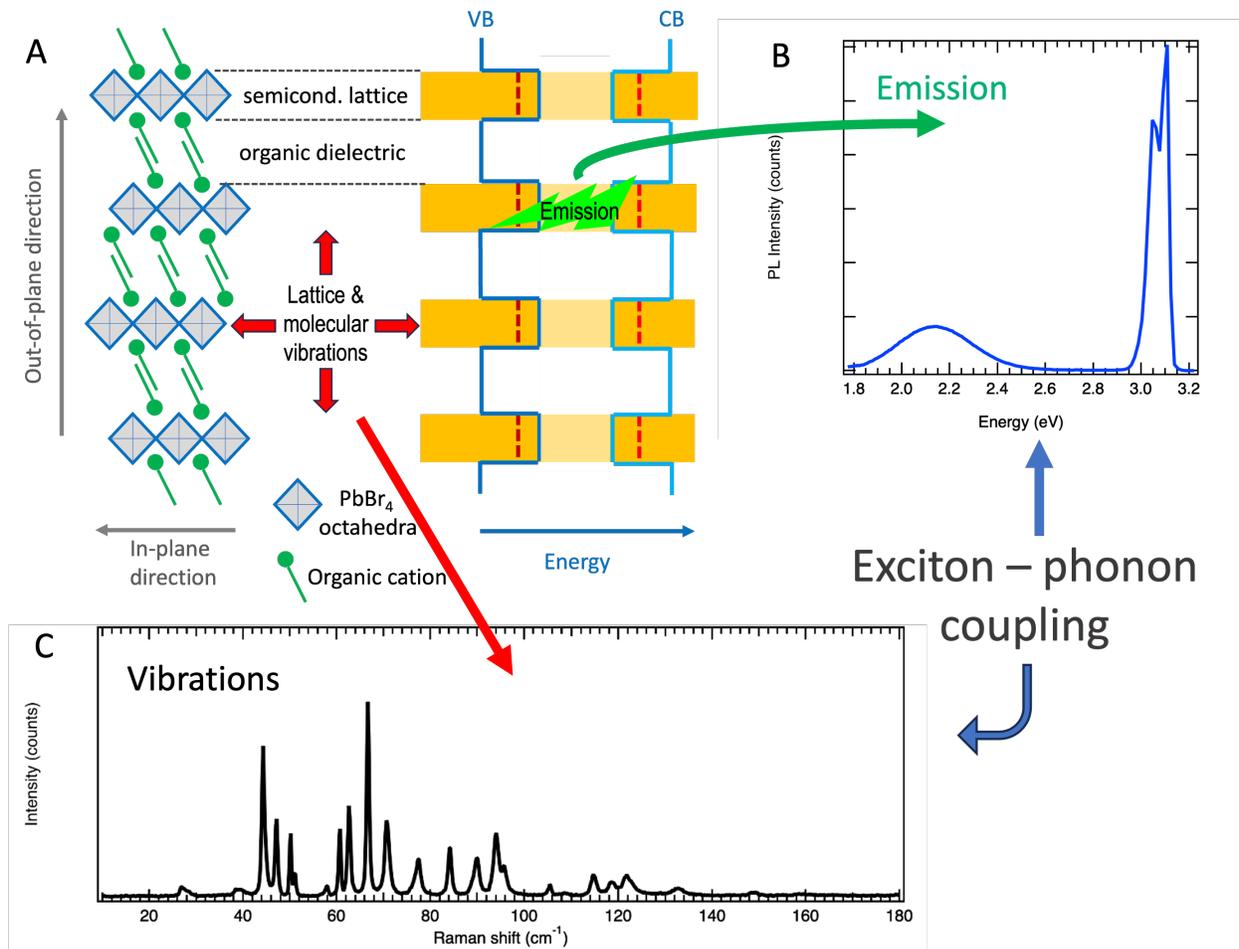

Figure 1. Two-dimensional metal-halide perovskites: architecture and typical emission and vibration spectra. A) Schematic illustration of the organic-inorganic layered structure that results in a periodic lattice of quantum wells in the vertical direction. B-C) Emission and Raman spectra recorded at cryogenic temperatures from single exfoliated $(UDA)_2PbBr_4$ flakes. Data adapted from ref. [20]

## 2. Two-dimensional layered perovskites: emission and vibration properties

The layered 2D perovskite architecture resembles stacks of semiconductor material quantum wells separated by dielectric layers formed by the organic molecules, as illustrated in Figure 1. This very sophisticated structure has many particular features: (i) it is strongly anisotropic, with quasi infinite in-plane extension and strong vertical confinement of the order of 1 nm, (ii) it combines the elastic properties of the inorganic octahedra lattice (that can be described in terms of acoustic and optical phonons) with those of the organic layer, where vibrations are originating from the molecular bonds, (iii) the overall lattice structure is relatively soft, therefore deformations can be easily induced by steric hindrance or electrical charges, (iv) the energy levels and electronic band structure depend both on the composition of the metal halide lattice and the species of organic cations.

Accordingly, such materials feature much more complex optical and elastic spectra compared to purely inorganic solid-state crystals, as shown in Figure 1B-C. The emission spectrum (Figure 1B) typically consists of a set of several peaks stemming from free excitons at the band edge, where the energy level structure depends on spatially diverse confinement effects, [25,26] and lattice asymmetries caused by distortions.[27] Furthermore, often a broad emission band at lower energy is observed that is related to self-trapped excitons (STEs) or defects.[28, 29] [30] Concerning the elastic properties, 2D layered perovskites feature a very large number of vibrational modes or phonon bands, as shown in the Raman spectrum in Figure 1C. These modes can be associated different kinds of oscillations, such as octahedra rocking or twisting, metal-halide bond bending or stretching, coupled modes organic head groups with the metal-halide octahedra, etc, as discussed in detail in a variety of works. [24, 31, 32] In the radiative recombination process, electron-phonon coupling plays an important role as it provide the necessary momentum and energy for the relaxation of the charge carriers to band-edge states with high transition dipole moments and oscillator strength. [33] Furthermore, the deformation of the lattice induced by the electrical charges can enable the self-trapping of excitons. [18, 34] From the spectra in Figure 1 we can easily see that electron-phonon coupling models that consider only one (typically LO) phonon mode will fall short of describing the response of 2DLPs.

**2.1 Impact of architecture anisotropy on the emission color**

In our work in Adv. Mater. in 2019, [35] we observed the tunability of the emission of ensembles of (DA)$_2$PbBr$_4$ microcrystals by applying pressures in the MPa range. This led to quenching of the low energy emission band under compression resulting in blue light emission, and white light when pressure was released (Figure 2A-B). Since compressive stress induced by pressure in the MPa range is not likely to result in significant lattice distortions,[36] we attributed this emission color change to the reduced thickness of the 2DLP film that goes along with horizontal alignment of the microcrystal plates. Both effects should result in reducing the reabsorption within the film, which enhances the contribution of the blue band edge emission. This interpretation was supported by our density functional theory (DFT) calculations of the absorption bands in the in-plane and out-of-plane directions of the octahedra lattice (see Figure 2C-D). We further explored this concept by embedding (BzA)$_2$PbBr$_4$ microcrystals in a stretchable PDMS film that enabled mechanical switching of the emission intensity by stretch and release of the free-standing film, [37] as shown in Figure 2E. Also in this case the re-absorption was reduced in the stretched film, which lead to higher emission intensity and spectral shifts of the emission peak.

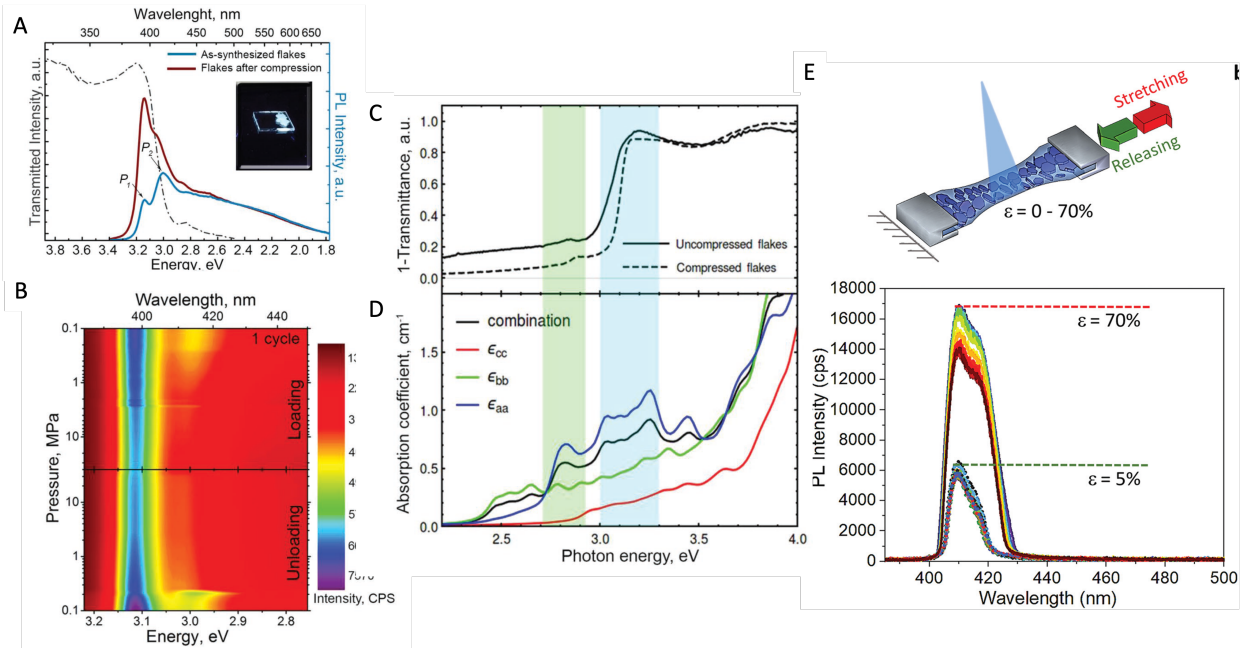

Figure 2. Emission tunability of 2DLP ensembles by stress induced alignment of the microcrystals. A-B) Change of the PL spectrum under compression of the ensemble by applying pressures in the MPa range. (C) Calculated transmittance of the microcrystal ensemble with and without compression. D) Simulated absorption along the crystallographic directions. Reproduced with permission from ref. [35]. E) Illustration of the stretchable PDMS film loaded with 2DLP microcrystals, and emission spectra for various cycles under 5% and 70% stretch. Reproduced with permission from ref. [37].

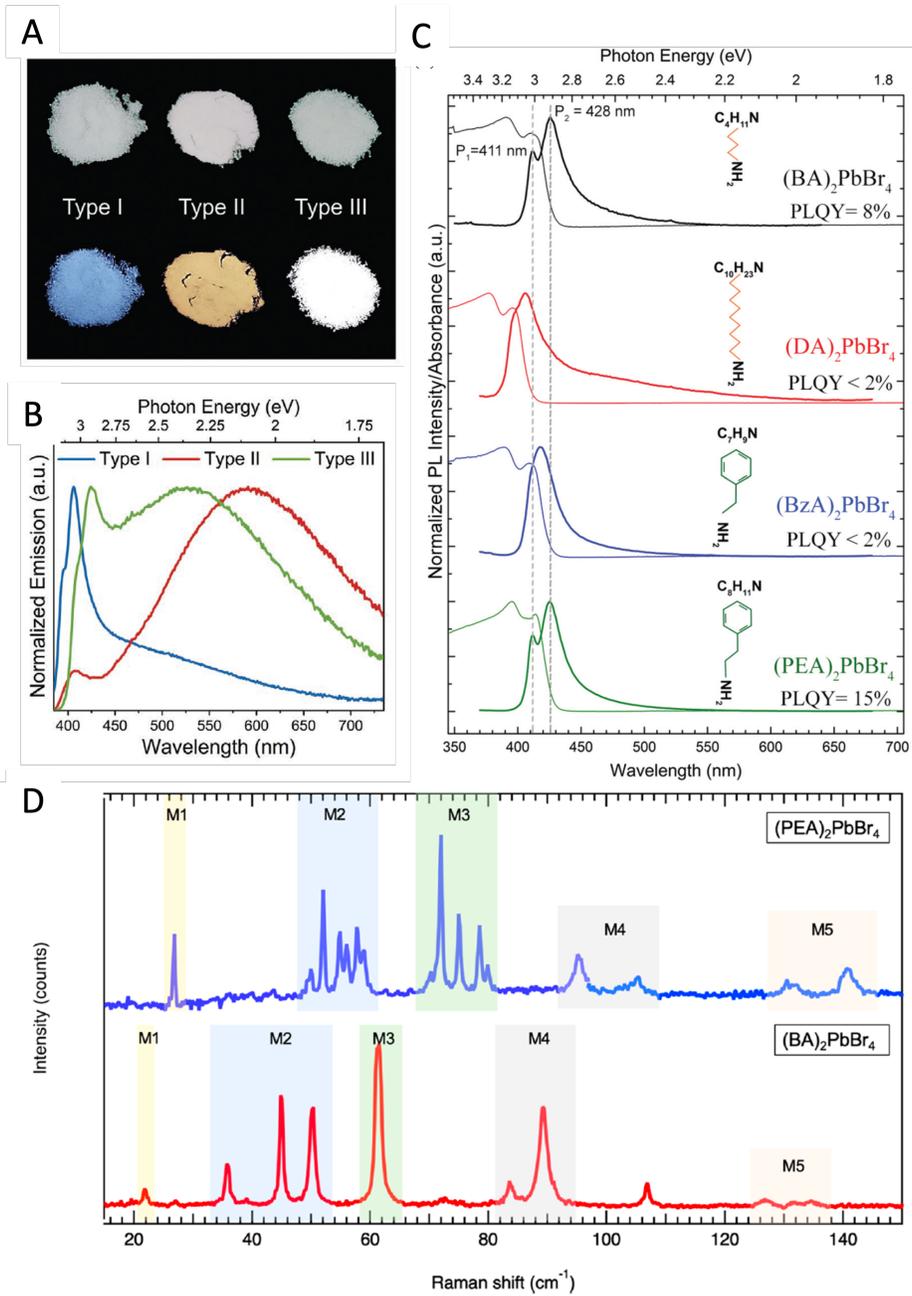

Figure 3. Exciton and phonon bands of 2D lead-bromide perovskites with different organic phases. A-B) 2D perovskite powders with organic cations that have different amine bonding groups under day (top) and UV (bottom) light showing different colors (A) and emission spectra (B). Reproduced with permission from ref. [29]. C) Emission and absorption spectra recorded from powders of 2D perovskites with primary amine bonding groups (reproduced with permission from ref. [38]). D) Raman spectra of 2D perovskites with different primary amine cations recorded at cryogenic temperatures, where phonon bands attributed to similar vibration modes are highlighted by the shaded background (reproduced with permission from ref. [24]).

**2.2 Impact of organic cations on emission color and phonon bands**

Figure 3 shows the emission spectra of various lead-bromide 2DLPs that differ in their organic cations. In our work in ref. [29], we classified the cations according to their binding head groups in type-I, type-II and type-III ammoniums. Figure 3A-B shows that the emission color depends strongly on the ammonium type, allowing to tune from almost pure (blue) band edge emission to warm and cold white emission that is dominated by (self-) trapped excitons. Within the primary (type-I) ammonium 2DLPs, we investigated molecules with aliphatic chain and phenyl-ring backbones and found slight shifts in the band edge emission wavelength that could be mainly attributed to different dielectric confinement, as shown in Figure 3C. Furthermore, the photoluminescence (PL) quantum yield varied strongly with the choice of organic cation, which could be related to differences in defects and electron-phonon coupling. The intensity of the broad band emission associated to self-trapped excitons (STE) depended strongly on the type of head group, as evident in Figure 3B. Modeling of the NH-Br bonds of the head group in the voids of the octahedra lattice evidenced the conformation of the organic cations and the induced lattice distortions. The more bulky head groups with an additional $CH_3$ group led to increased STE emission, which was related to Jahn-Teller distortions.[39] For the evaluation of the electron-phonon coupling strength related to such broad emission, a model for the line width broadening of the emission of color centers[40] has been successfully applied to the thermal emission behavior of double perovskites. [15, 41] However, when we applied this analysis to the 2DPLs with type II and type III ammonium molecules, we did not obtain a satisfactory fitting of the emission line width, which we attributed to the large number of dominant phonon resonances that we observed in these materials.[20]

With respect to complex vibrational properties of the 2DLPs, our low-frequency Raman spectroscopy studies (Figure 3D) revealed very different phonon bands for $(PEA)_2PbBr_4$ (phenyl-ring molecule with more rigid $\pi$-$\pi$ stacking in the Van-der-Waals gap) than for $(BA)_2PbBr_4$ that has a tail-to-tail arrangement of the aliphatic chains. In our work in ref. [24] we assigned different sets of Raman peaks to phonon bands that we associated to distinct vibrational motions (see Table 1), in agreement with other works.[31] The lowest frequency band M1 was correlated to octahedra twisting and rocking, the M2 band to stretching and bending motions involving the (heavy) Pb ions in the metal halide lattice, and the M3 band to motions dominated by the lighter Br ions. The higher frequency bands M4 and M5 were associated to out-of-plane vibrations involving the organic cations. The phonon bands of the PEA system are at higher frequencies than those of the BA system, which can be attributed to the higher rigidity of the phenyl ring system. Interestingly, we found more phonon modes in the $(PEA)_2PbBr_4$ bands (with respect to $(BA)_2PbBr_4$), which we tentatively assigned to the mode splitting due to vibrational coupling between the layers. In ref. [20] we also explored the interpretation of coupled acoustic modes for the origin of the low-frequency Raman peaks by applying the Rytov model of backfolded superlattice modes,[42] however, we could only achieve reasonable agreement with our data by assuming elastic parameters for the organic-inorganic layers that were different from those of their bulk material components.

Table 1. Frequencies of the vibrational modes discerned in the Raman spectra for (PEA)$_2$PbBr$_4$ and (BA)$_2$PbBr$_4$ flakes at T=4K, together with their symmetry, and the tentative assignment of the dominant vibrational motion. Reproduced with permission from ref. [24]

| Band | (PEA)$_2$PbBr$_4$ freq. (cm$^{-1}$) | (BA)$_2$PbBr$_4$ freq. (cm$^{-1}$) | Irreducible representation (D$_{2h}$ symmetry) | Vibrational motion |
|---|---|---|---|---|
| M1 | 26.8 | 21.8 | B$_{1g}$, B$_{3g}$ | Octahedra rocking/twisting |
| M2 | 52.4 | 35.7 | A$_g$ | Pb-Br bond bending |
| | 54.9 | | | |
| | 56.0 | 44.9 | | |
| | 57.7 | 50.3 | | |
| | 58.8 | | | |
| M3 | 70.3 | 61.5 | B$_{2g}$ | Pb-Br bond bending and twisting; Br-Pb-Br scissoring in the octahedral plane |
| | 72.0 | | | |
| | 75.0 | | | |
| | 78.5 | | | |
| | 80 | | | |
| M4 | 95.1 | 83.6 | A$_g$ | Out-of-plane Pb-Br bond stretching |
| | 105.3 | 89.1 | | |
| M5 | 131.6 | 106.8 | A$_g$ | In/Out-of-plane Pb-Br bond stretching |
| | 140.7 | 132 | | |

### 3. Phonon directionality and emission polarization of single microcrystals

From our studies on ensembles (powders) of 2DPL microcrystals we observed that the choice of the organic cation has a strong impact both on the optical and vibrational properties of the material. However, such powder measurements are not adequate to explore the features that originate from the strong anisotropy of the 2DLP microcrystals, since they average over many different orientations of the individual crystals. We therefore turned to investigate single microcrystals obtained by mechanical (scotch-tape) exfoliation (Figure 4A). Such exfoliation led in many cases to flakes with regular edges (straight, and with sharp defined angles at the corners) that gave access to explore the effect of the orientation of the octahedra lattice in single crystalline regions with micro-PL and micro-Raman spectroscopy. Mechanical exfoliation results in microcrystal flakes with thicknesses ranging from few nanometers up to several hundreds of nanometers, which leads to the different colors of the flakes in the optical microscope image in Figure 4A, due to the thin film optical interference effect. The thinnest flakes that we measured with this method were 8-10 nm in height (Figure 4B and ref.[38]), which, with interlayer spacing of ca 1.5 nm, corresponds to approximately six monolayers of octahedra lattice planes. Interestingly, in ref [38] we observed that for material from the same batch, the PL from powders had a double peak structure in the band edge emission, while from the exfoliated flakes we measured only a single peak that coincides with the higher energy peak of the powders (Figure 4C). Furthermore, the PL quantum yield from the exfoliated flakes was much higher (42%) than that of the powders (15%). We attributed this effect to the different confinement in out-of-plane and in-plane directions (as discussed in Figure 2D), that in ensembles of differently oriented microcrystals favors reabsorption of the higher energy light and re-emission from lower energy levels with lower transition rates. This interpretation was supported by the longer decay time of the low-energy emission peak, [38] and by modeling the absorption coefficients in the different directions [43].

With our low-frequency Raman spectroscopy experiments in ref. [24], we discovered that the phonon spectra measured from single crystalline regions of lead-bromide microcrystals depend strongly on the orientation of the microcrystal with respect to polarization optics of the setup, as shown in Figure 4D-E. Comparing spectra recorded with parallel (horizontal) and diagonal configuration reveals that some phonon bands appear only at certain angles, thus their intensity is highly directional (see for example the band around 50 cm$^{-1}$ in Figure 4D-E that is absent in the spectra of the diagonal flake). In this work we investigated also flakes with different thickness and found that the profile of the Raman spectra was not affected by the number of layers in the flake, only the signal scaled with the overall thickness. However, the band edge PL peak showed a slight dependence on the flake thickness, manifesting a red shift of few nm with increasing number of layers from 15 to 75 (Figure S4 in ref. [24]), which we tentatively assigned to self-absorption.

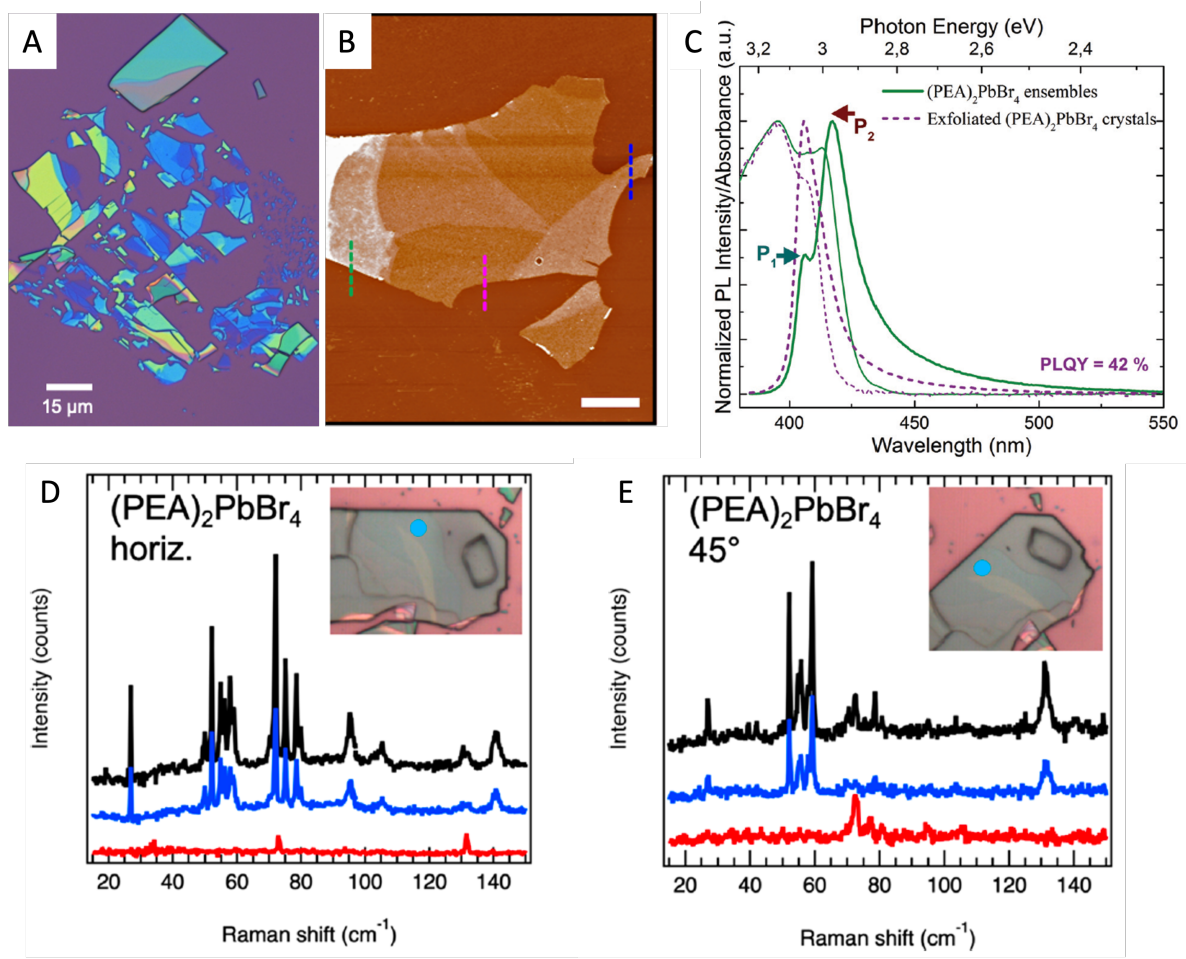

Figure 4. Exfoliation of single 2D perovskite flakes and directional dependence of the optical and vibrational spectra. Optical (A) and atomic force (B) microscopy images of exfoliated flakes where the height profiles along the green, pink, and blue dashed lines reveal thicknesses of 20, 14 and 10 nm, respectively. C) Photoluminescence (PL) spectra recorded from powders (ensembles) and exfoliated crystals. Reproduced with permission from ref. [38]. D-E) Raman spectra recorded from a single flake with horizontal (D) or diagonal) orientation with respect to the detection optics. The blue dot indicates the excitation/detection spot. Black: unpolarized, blue: polarized, red: depolarized. Reproduced with permission from ref [24].

To delve deeper into the directionality of the different phonon bands, we extend our Raman setup to angle-resolved polarized Raman (ARPR) spectroscopy. This was implemented by inserting a $\lambda/2$ wave plate in the common optical excitation and detection path (above the microscope objective), which allowed to rotate the linear polarization of the light with respect to the sample. Here the rotation of an angle of $\theta/2$ of the wave plate is equivalent to a rotation of $-\theta$ of the sample. [20, 44]. With this setup we recorded the angular dependence of the Raman signal in polarized (VV) and depolarized (HV) configurations, as illustrated in Figure 5 A-B. Colormaps of the Raman intensity for versus angle and frequency are displayed in Figure 5 C-D (recorded from $(BA)_2PbBr_4$ at T=5K with below band gap excitation at 633 nm). Clearly, the different phonon modes show very diverse angular behavior, featuring one or two intensity maxima, or almost constant intensity, in agreement with ref. [19] Note that the maxima of the same mode are shifted by 45° in HV with respect to VV, which indicates directionality along the two orthogonal axes that we associated with the major axes of the octahedra lattice. Such directional behavior can be rationalized and modeled by different symmetries of the Raman tensor, as shown by the polar plots in Figure 5E.

For non-resonant excitation, the Raman scattering intensity can be expressed by the Raman tensor as:

$$I_R \propto \left| \boldsymbol{e}_s^T \cdot \vec{R} \cdot \boldsymbol{e}_L \right|^2 \quad (1)$$

Here $\boldsymbol{e}_L$ and $\boldsymbol{e}_s$ are the vectors of the incident and scattered light, $\boldsymbol{e}_L = \boldsymbol{e}_s = (\cos\theta, \sin\theta, 0)$ for VV configuration ($\theta$ is the relative angle between laser polarization vector and the crystallographic axis that was controlled by the $\lambda/2$ wave plate), while $\boldsymbol{e}_L = (\cos\theta, \sin\theta, 0), \boldsymbol{e}_s = (\sin\theta, -\cos\theta, 0)$ for HV configuration, and $\vec{R}$ is the Raman tensor with

$$\text{With } \vec{R} = \begin{pmatrix} a & d & \cdot \\ d & b & \cdot \\ \cdot & \cdot & \cdot \end{pmatrix}, \quad (2)$$

where the coefficients related to the z-direction are replaced by a dot, since in our in-plane analysis they are not considered. Under VV polarization, the ARPR intensity of one phonon follows $I \propto |a\cos^2\theta + b\sin^2\theta + 2d\sin\theta\cos\theta|^2$; while under HV polarization, it is $I \propto |(a-b)\sin\theta\cos\theta + d|^2$. By comparing the experimental and calculated angular patterns, it is possible to assign the phonon symmetries. For example, the isotropic $A_{1g}$- like mode for a=b=1 and d=0, and $T_{2g}$ and $E_g$ like symmetries for the quadrupolar modes.

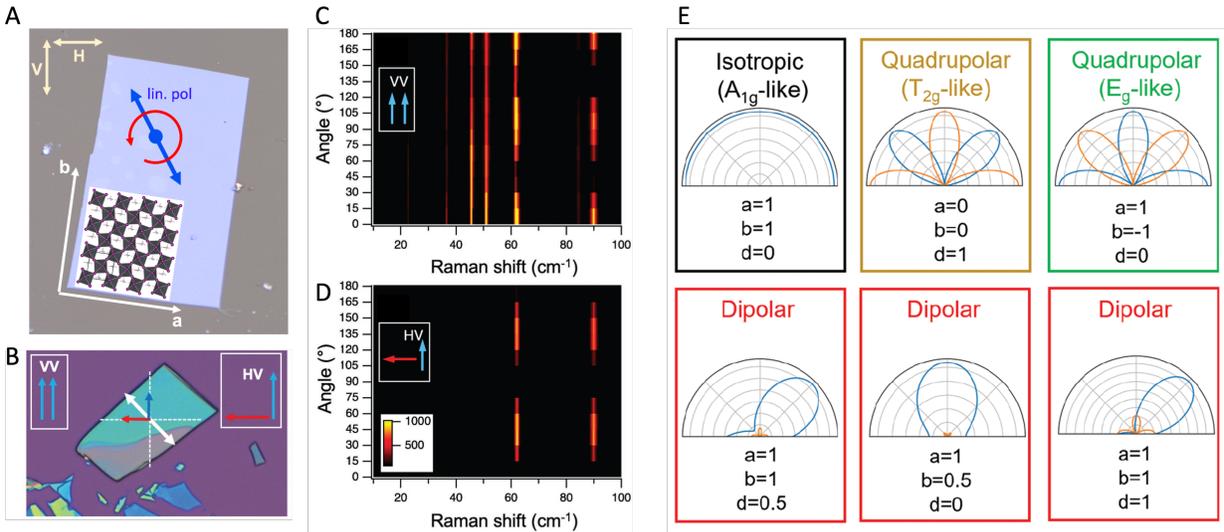

Figure 5. Angle-resolved polarized Raman spectroscopy on single flakes. A) Illustration of the relative orientation of the perovskite octahedra lattice with respect to the polarizers in the excitation/detection paths (V,H), and the rotation of the linear polarization with respect to the sample. C-D) Color map of angle-resolved Raman spectra recorded in polarized (V,V) and depolarized (H,V) configurations. Reproduced with permission from ref. [20]. E) Relation of the Raman tensor to the angular mode intensities, resulting in isotropic, bipolar or quadrupolar modes.

We also obtained detailed insights into the directionality of the vibrational modes by a careful geometrical analysis on the angle-resolved Raman spectra in polarized (VV) and depolarized (HV) configurations, as specified in Figure 6 A-C. We could identify modes along both major axes of the octahedra lattice (quadrupolar) that can stem from octahedra rocking, or metal halide bond stretching (dark blue), modes along only one direction (dipolar), with oscillations either along the same direction (green – bond stretching) or perpendicular to it (red- bond bending), or both (yellow-bond scissoring), and isotropic modes that we assigned to out of plane oscillations. Interestingly, for the UDA sample in Figure 6A we found strong contributions of quadrupolar modes along directions diagonal to the other set of quadrupolar modes, which has a significant impact on the linear polarization of the emission, as we found out in our recent study that enabled us to record also angle-resolved polarized emission in addition to the Raman signal. [22] Our works on two-dimensional double perovskites (see Figure 6D) and on lead-iodide 2DLPs with different organic cations showed that such highly complex angular behavior of the vibrational bands is not restricted to the lead bromide system and should be a general consequence of the sophisticated anisotropic architecture of 2DLPs.

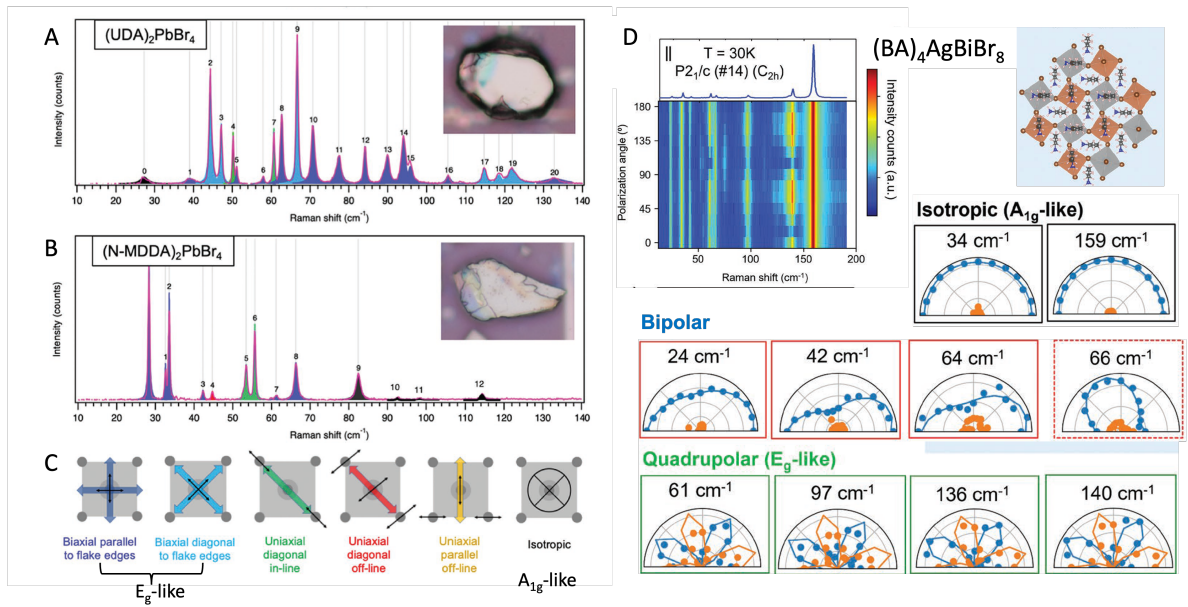

Figure 6. Assignment of the directional behavior of the phonon modes detected by Raman spectroscopy from single flakes. A-C) Raman spectra recorded from lead-bromide 2D perovskites at cryogenic temperature, the directional properties color coded in A-B) and illustrated in C) were extracted from angular intensity mapping in polarized and depolarized configurations. Reproduced with permission from ref. [20]. D) Angle-resolved Raman data from 2D double perovskites and classification of the modes with respect to the symmetries of the Raman tensor. Reproduced with permission from ref. [23]

**3.1 Combined emission and Raman spectroscopy on single microcrystals**

Concerning the impact of electron-phonon coupling to the microcrystal emission, it would be very interesting to investigate the same microcrystal both with angle-resolved Raman and PL spectroscopy. Unfortunately, the optics of our cryogenic micro-PL (micro-Raman) setup are not UV compatible, and therefore excitation far above the band gap of the lead-bromide 2DLPs was not possible at cryogenic temperatures. However, the exciton fine structure only emerges below ca 100K, as will be shown later. Therefore, we turned to lead-iodide 2DLPS that emit in the green and where blue laser light can be used for the PL and Raman studies.[22] Our results for BA and UDA ammonium molecules as organic cations are displayed in Figure 7. Interestingly, although the difference of these molecules is only in their aliphatic chain length (4 carons versus 11 carbons), they feature very different emission and vibrational properties, both in their mode spectrum and in their angular dependence. For $(BA)_2PbI_4$ (Figure 7A) we observe two emission modes with orthogonal in-plane polarization (in agreement with ref. [45]), with goes along with dominant Raman bands around 25 cm$^{-1}$ and 50 cm$^{-1}$ that under resonant (above band gap) excitation have their intensity maxima at the same angles as the PL. [22] With non-resonant excitation (Figure 7B) we can resolve a more detailed phonon spectrum, where also few weaker side bands with their maxima shifted by 45 ° can be noted, for example in the range from 50-60 cm$^{-1}$.

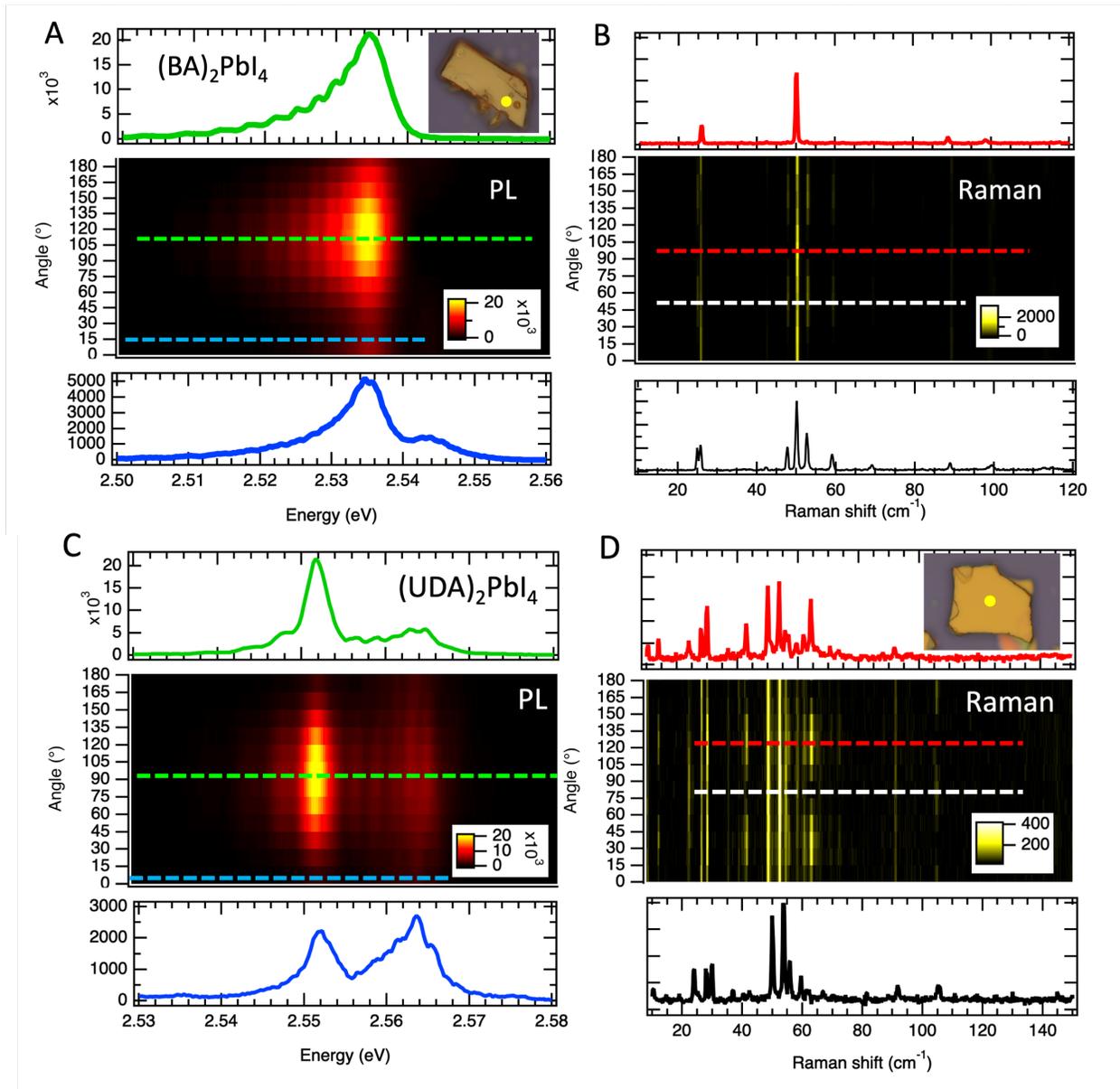

Figure 7. Correlation of PL and Raman spectra recorded of the same lead-iodide microcrystal. Insets show optical microscope images of the investigated flakes and the yellow dots indicate the excitation/detection spot. A,C) Angle resolved PL spectra from single exfoliated $(BA)_2PbI_4$ and $(UDA)_2PbI_4$ microcrystals. Spectra at the angles indicated by the green (blue) dashed lines are shown in the top (bottom) panels. B, D) Angle dependence of the phonon bands measured by non-resonant Raman spectroscopy, the bottom and top panels shows spectra recorded at the angles indicated by the red(white) dashed lines. Modified from ref. [22].

Turning to (UDA)$_2$PbI$_4$, we observe a very different behavior. Here all PL peaks have their intensity maximum at the same polarization angle (at 90° for this flake), and the intensity of the phonon modes under resonant excitation is almost independent of the polarization angle. Concerning the different PL peaks, we can distinguish certain set of peaks that have different intensity increases with the angle, therefore a signature of the in-plane polarization of the confined modes is present. With non-resonant Raman scattering (Figure 7C) we distinguish a very large number of phonon modes in the low frequency regime that can be assigned in analogy to the lead-bromide modes discussed above. The particular feature of the Raman bands of (UDA)$_2$PbI$_4$, is that a large number of modes with polarization in the diagonal directions is present. These non-orthogonal modes can couple, and therefore phonon modes with different directionality can contribute to the electron-phonon coupling related to a specific emission peak. This mixes modes with different directionality, and thus all emission peaks have their maximum at the same angle. We note that this angle is shifted with respect to the directions of the phonon modes.

**4. Temperature dependence of emission and electron-phonon coupling**

The strong impact of the low-frequency phonons and electron-phonon coupling on the emission properties of the 2DLPs is also evident in the temperature dependence of the photoluminescence. The different sets of peaks that characterize the exciton fine structure emerge at temperatures below 100 K (Figure 8), thus in a range that corresponds to thermal energies below 10 meV, in which we observe the large variety of low-frequency phonon bands. In our work in ref. [21] we investigated the temperature dependence of the emission of lead-bromide 2DLPs with different organic cations and observed different sets of emission peaks for the different materials that emerged below 100K, as shown in Figure 8A. To evaluate the electron-phonon coupling we focused on the high energy peak in the band edge emission that is the most dominant in this temperature range, and analyzed its thermal broadening by Gaussian line width fitting.

The thermal broadening of the PL of polar semiconductors can be described by: [17]

$$\Gamma(T) = \Gamma_0 + \Gamma_{LO}/[\exp(E_{LO}/k_B T) - 1] \quad (3)$$

if we neglect coupling to acoustic phonons[46] and the scattering due to impurities. Here $\Gamma_0$ is the PL line width at zero temperature, $\Gamma_{LO}$ the coefficient for the exciton-optical phonon interaction, $E_{LO}$ the average energy of the longitudinal optical (LO) phonon modes involved in the coupling, and $k_B$ is the Boltzmann constant. As can be seen in panels A (b-d) in Figure 8, the fitting works quite well for the BA system that has relatively few phonon modes in the low-frequency range, and yields an LO phonon energy of 8.3 meV that matches relatively well the M3 band discussed in Figure 3. For the UDA and N-MDDA systems the fitting is much less accurate (precise fitting was also made difficult by strong changes in the spectral shape with temperature), and we obtained high values for the phonon energy of 26 meV and 18 meV that did not correspond to dominant Raman bands in our experiments. Furthermore, we obtained a very high value of 360 meV for the electron-phonon coupling, which goes along with the very large number of low-frequency phonon modes that we observed for the UDA system. We therefore hypothesized that the combination of different phonon modes and two-phonon scattering (as reported for bilayer $WSe_2$.[47]) could be the reason for obtaining such high phonon energy values in the fitting. However, this interpretation was based on data from non-resonant Raman scattering, and phonon mode intensities can strongly change with resonance conditions. Furthermore, some weak phonon bands at energies of close to 30 meV were present in the non-resonant spectra of $(UDA)_2PbBr_4$. This motivated the investigation of single 2DLP flakes under resonant and non-resonant conditions in combination with the measurement of their angular emission properties [22] that we already discussed in Figure 7. In these experiments we did not observe the appearance of strong phonon bands at high energies (in the 20-40 meV range), but we found evidence of the mixing of phonon bands with different directionality as discussed above, which supports the interpretation of the involvement of two-phonon scattering events in the thermal emission line width broadening. The temperature dependent emission spectra of $(BA)_2PbI_4$ and $(UDA)_2PbI_4$ in the range below 100K are shown in Figure 8 B-C. Gaussian fitting to extract the line with is hindered by the change in spectral shape, since for $(BA)_2PbI_4$ a pronounced low energy tail emerges with decreasing temperature, while for $(UDA)_2PbI_4$ the emission band splits into high and low energy doublet. This behavior evidences the thermal (phonon) coupling of the different exciton levels at the band edge, in which the highest energy peak that can be related to the out-of-plane confinement has the strongest oscillator strength, and therefore dominates the emission.

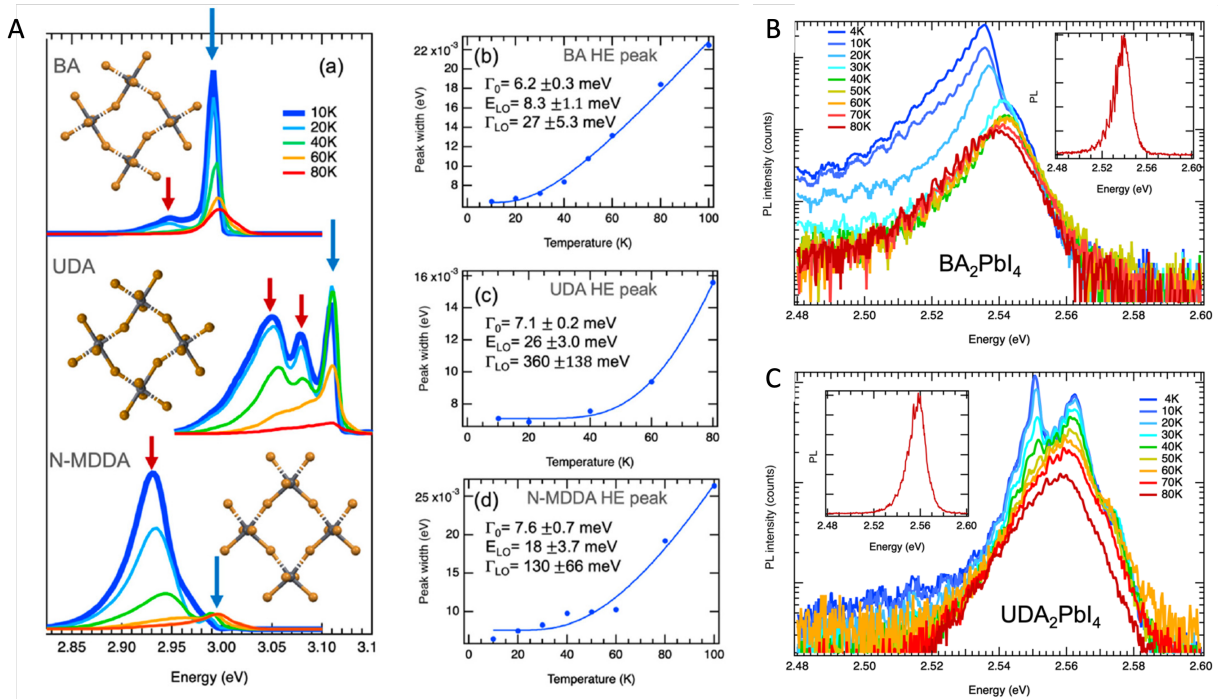

Figure 8. Temperature dependence of the band edge emission. A) (a) Photoluminescence (PL) recorded from 2D lead-bromide microcrystal powders with BA, UDA and N-MDDA as organic cations at cryogenic temperatures with laser excitation at 375 nm. (b-d) Fitting of the Gaussian line width broadening of the peak indicated by the blue arrow with equation (3). Reproduced with permission from ref. [21] B-C) PL spectra recorded from single lead-iodide microcrystals shown on a semi-logarithmic scale. Reproduced with permission from ref. [22].

## 5. Summary and outlook

The emission and Raman spectroscopy studies showed that two-dimensional hybrid organic-inorganic metal halide crystals are a very particular system in which the properties of soft organic phases and rigid inorganic lattices are strongly intertwined. The inorganic (semiconducting) metal halide octahedra lattice is at the basis of the electronic level structure, but the organic phases alter the optoelectronic properties due to lattice distortions, different confinement conditions, and diverse vibrational coupling. This on one hand allows for a very large freedom to engineer materials with diverse properties, but on the other hand also complicates the modeling and prediction of their optoelectronic behavior. Furthermore, due to their layered architecture these materials feature strong anisotropies in their optical, electrical and elastic properties, and our single microcrystal studies evidenced that in addition to the obvious out-of-plane anisotropies, there are also subtle in-plane effects that affect the phonon bands, electron-phonon coupling and thereby the emission polarization. Effects of the architecture anisotropy become particularly evident when single microcrystals are at the center of the experiment or device. For example, the reabsorption of the emitted light is favored in ensembles of differently oriented microcrystals (due to the band gap differences along the different crystal directions), while single microcrystals and thin film of aligned crystals show much higher emission quantum yield. Furthermore, the phonon spectra of single microcrystals depend strongly on the relative orientation of the microcrystal lattice with the polarization of the incident light, because many phonon bands feature a strong directionality. Also the emission manifests polarization anisotropies, that in some cases go along with the spatial confinement that leads to level splitting in the exciton fine structure, and therefore polarization along the corresponding x, y, z spatial directions.[45] However, in 2DLPs that feature phonon bands in non-orthogonal directions, in particular along major crystal axes and their diagonals, multi-phonon coupling in the optical relaxation process can mix the polarization dependence and alter the emission polarization. The involvement of several phonon modes in the emission process renders commonly used models for electron phonon coupling (based on coupling to a single LO phonon mode) inaccurate. On the other hand, a deeper understanding of the electron phonon coupling in 2DLPs can open avenues to tailor their emission properties of 2DLPs by acting on the type of organic cation. We saw on the example of BA and UDA that only a small change in the length of the aliphatic chain can significantly change the emission spectrum, polarization, and quantum yield. Such tunability can be highly interesting towards optical devices based on single microcrystals. Due to their high dielectric permittivity in the visible[48], 2DLP microcrystals can act as photonic cavities[49] and for example, give rise exciton-polaritons.[50] Here tuning the emission polarization will be highly interesting to enhance the performance and functionality of such single crystal photonic structures. Therefore, gaining knowledge on how the vibrational properties and exciton-phonon coupling determine the emission properties of 2DLPs will be highly versatile for optical applications of these materials, and we hope that our Account could give a useful contribution in this respect.

## AUTHOR INFORMATION

**Corresponding Author**

*roman.krahne@iit.it**Author Contributions**

The manuscript was written through contributions of all authors. All authors have given approval to the final version of the manuscript.

## ACKNOWLEDGMENT

R.K acknowledges funding by the European Union under Project 101131111 – DELIGHT. P.H. and M.L. acknowledge the support from National Natural Science Foundation of China (Grant Nos. 12322401 and 12127807), Beijing Nova Program (Grant No. 20230484301), Youth Innovation Promotion Association, Chinese Academy of Sciences (No. 2023125).**Selected References**

Dhanabalan, B.; Leng, Y.-C.; Biffi, G.; Lin, M.-L.; Tan, P.-H.; Infante, I.; Manna, L.; Arciniegas, M. P.; Krahne, R. Directional Anisotropy of the Vibrational Modes in 2D-Layered Perovskites. *ACS Nano* **2020**, *14*, 4689-4697

Lin, M.-L.; Dhanabalan, B.; Biffi, G.; Leng, Y.-C.; Kutkan, S.; Arciniegas, M. P.; Tan, P.-H.; Krahne, R. Correlating Symmetries of Low-Frequency Vibrations and Self-Trapped Excitons in Layered Perovskites for Light Emission with Different Colors. *Small* **2022**, *18*, 2106759

Martín-García, B.; Spirito, D.; Lin, M.-L.; Leng, Y.-C.; Artyukhin, S.; Tan, P.-H.; Krahne, R. Low-Frequency Phonon Modes in Layered Silver-Bismuth Double Perovskites: Symmetry, Polarity, and Relation to Phase Transitions (Advanced Optical Materials 14/2022). *Adv. Opt. Mater.* **2022**, *10*, 2270056

Krahne, R.; Schleusener, A.; Faraji, M.; Li, L.-H.; Lin, M.-L.; Tan, P.-H. Phonon Directionality Determines the Polarization of the Band-Edge Exciton Emission in Two-Dimensional Metal Halide Perovskites. *arXiv preprint arXiv:2404.11367* **2024**

SYNOPSIS (Word Style "SN_Synopsis_TOC").

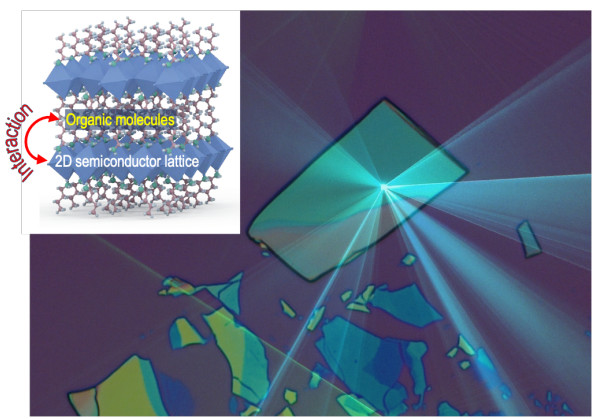